# Dissipation in quantum tunnel junctions

Edgar J. Patiño, L. Rios E., N. G. Kelkar and Daniel Lopez

*Departamento de Física, Superconductivity and Nanodevices Laboratory, Universidad de los Andes, Carrera 1 No. 18A 12, Bogotá, Colombia*

Based on experimental data, we propose a model to evaluate the energy dissipated during quantum tunneling processes in solid-state junctions. This model incorporates a nonlinear friction force expressed in the general form $f(x) = \gamma v(x)^\alpha$, where $\gamma$ is the frictional coefficient, which is fitted to data. We study this by applying voltages just below the barrier height up to near break down voltages. Furthermore, by lowering the temperature and adjusting the applied voltage to the junction, the effect on dissipation caused by the variation in barrier height is examined. We underline that the crucial dependency of dissipation on the fraction of particle energy lost is modulated by two primary mechanisms: the application of voltage and the variation of temperature. The fraction of energy dissipated decreases in general for increasing energies of the tunneling particles at a given temperature. However, for a given energy of the tunneling particle, the present work demonstrates a turning point at a temperature of 137 K, after which the dissipated energy starts increasing for higher temperatures. The latter can possibly be due to the increase of electron-phonon interactions which become predominant over barrier height reduction at higher temperatures and hence we identify T = 137 K as a critical temperature for change in dissipative characteristics of the solid-state junction under consideration. Notably, also the study identifies significant changes in dissipation parameters, γ and α, above 137 K, exhibiting a linear decline and underscoring the importance of further research at higher temperatures.

*Corresponding author: epatino@uniandes.edu.co

## I. INTRODUCTION

In quantum mechanics, the principle of energy conservation holds during the tunneling process. If particles do not interact with the environment (such as in vacuum in STM microscopy) the tunneling process occurs without gaining or loosing energy.

On the other hand, when tunneling occurs in a dissipative environment such as within a solid-state device where electrons interact with lattice vibrations (phonons), there is coupling between tunneling particles and the environment. This interaction can lead to dissipation such as the described in Caldeira-Leggett model. This coupling can result in the loss of phase coherence and energy exchange with the environment, affecting the tunneling dynamics. For example, magnetic tunnel junctions have attracted many research efforts due to potential applications in quantum computing and information storage systems [1,2]. These junctions comprise of two magnetic electrodes separated by an ultra-thin insulating film, which represents the potential barrier in the device. When an electron, subject to a voltage, crosses a tunneling barrier there are inelastic electron scattering processes that lead to dissipation of energy. Experimentally, this can be quantified by measuring the tunneling resistance. This energy at low voltages is expected to be dissipated in the form of heat, however, if the applied voltage exceeds approximately one volt, the junction emits infrared radiation, which arises from high frequency component of the current´s shot noise [3–5]. Given the fact that solid state devices are packed into ever smaller areas within computers and electronic devices, investigations of heat dissipation and its connection to the underlying electron-phonon interactions has gained importance [6].

In most tunnel junctions, the standard choice for the potential barrier is a film of $Al_2O_3$ or MgO of approximately 20 Å of thickness. Several theoretical models are available for the study of tunneling currents, where dissipation is so small that it can be neglected. However, as we will show, dissipation can no longer be neglected for large, applied voltages. This limitation confines most models to low voltages only. The few theoretical models that include dissipation in quantum processes have two main approaches: the first, where dissipation comes from the coupling of the system with a heat bath in a microscopic formulation [7]. The second one, with a semiclassical consideration that includes dissipation in a damped system through a frictional force [8]. The latter approach provides a simple logical model with a methodology that can be included in Simmons's current-voltage characteristics to study dissipation in tunnel junctions [9].

To learn more about dissipation and its significance, in the present work, we analyze current-voltage characteristics of large area $Al/Al_2O_3/Al$ tunnel junctions. Based on our experimental data, and on Simmons model with an additional consideration of a dissipative frictional force, we propose a dissipation model that works for the full voltage range up to the tunnel barrier height $\frac{\phi_o}{e}$.

To check the validity of this model we compare it with Simmons model [10] which, given its accuracy in modeling tunneling currents, at low applied voltages, is nowadays the most widely used model in the literature. However, as we will show at larger voltages, the model fails to accurately predict the experimental results in pristine Al/Al$_2$O$_3$/Al tunnel junctions even when image forces are included.

To our knowledge there is only one report on Al/Al$_2$O$_3$/Au (-Ag, -Cu) tunnel junctions [11], where a third electrode has been included to generate trapezoidal barriers. Here image forces are incorporated using Simmons model, however little information is provided on the quality of the shown fit.

Consequently, here we propose an extended model that includes dissipation of energy of the tunneling process and maintains the great accuracy of its predecessor.

Our approach is empirical as we look for the best fit that explains our data using the minimum number of free parameters, thereby obtaining valuable information about dissipation.

For a particle that tunnels across a potential barrier of width "s" and height "$\phi_o$" the amount of energy lost (or dissipated) while traversing the barrier along the x-direction can be expressed as $\Delta E(x)$. Previously two of the present authors of this publication, Kelkar and Patiño [9], proposed a model to explore dissipative tunneling across a rectangular barrier in the low voltage regime $eV \sim 0$ where the dissipation is small compared to the energy of the incident particle; $\Delta E(x) \ll \phi_o - E$.

In the present work, we extend this previous effort for higher dissipation in the intermediate voltage regime $eV \leq \phi_o$ where $\Delta E(x) < \phi_o - E$. We do this using a phenomenological approach to study the dissipation of tunnel junctions in the full voltage or energy range, provided that the dissipated energy is below the incident energy of the particle i.e. $\Delta E(x) < E$.

Previous experimental works at voltages close to the breakdown value [12–14] reported multiple issues. For instance, in [12] the presence of hotspots greatly affected the tunneling effective area. In [13], the breakdown voltages were lower than 1.3 V, not reaching the barrier height of around 2 eV. In [14], acceptable fittings to the current voltage characterization could only be made when multiple layers of oxide structure or multiple barriers were assumed.

On the other hand, in the present investigation, we present an improved version of our previous dissipation model [9], this time, including the full energy range where voltages are close to $\phi_o/e$.

## II. SAMPLE PREPARATION AND CHARACTERIZATION

Tunnel junctions with structure Al/Al$_2$O$_3$/Al were fabricated with a series of mechanical masks in a HV e-beam system. The structure was grown on a Si substrate with a 300 nm thick SiO oxide buffer layer to avoid current leaks from the sample to the substrate. Each layer of the structure was grown with a base pressure lower than $1\times10^{-5}$ Torr and evaporation pressure of around $1\times10^{-4}$ Torr. The fabrication of the oxide layer Al$_2$O$_3$ was achieved by exposing the first Al layer to a DC oxygen plasma of 25W, 500V, 50mA at $2.3\times10^{-2}$ Torr for 25 minutes. The tunneling area of the final device consists of a square of approximately 500×500 μm$^2$. This area allows the junction to withstand large tunneling currents without suffering severe damage and implies that the device has large enough capacitance so that there is no charge accumulation due to the Coulomb Blockade effect [15].

Indeed, the minimum energy needed to store charge $E_c = e^2/2C$, is inversely proportional to the capacitance C. This energy can be estimated from the expression $C = \varepsilon\frac{A}{d}$ where A is the tunneling area and d is the dielectric (Al$_2$O$_3$) thickness. Using the dielectric constant of Aluminum Oxide with a value between 9 and 12 $\varepsilon_0$, and an average dielectric thickness of 20 Å we obtain a value of C ~ $1\times10^{-8}$ F and an energy of $E_C \sim 1\times10^{-30}$ J is found. This is smaller by several orders of magnitude compared to the minimum thermal energy of the junction $k_bT \sim 5.5\times10^{-23}$ J (at 4K). Therefore, Coulomb Blockade is absent in our junctions. This is in clear contrast with the capacitance required to obtain single electron tunneling due to Coulomb blockade in previous works [15,16], where the values of capacitance are less than $1\times10^{-15}$ F.

Current-Voltage (I-V) characteristics were first taken at low-temperature, starting at zero current, and increasing up to 50 μA. The corresponding voltage for each current was measured in a four-terminal geometry. I-V characterizations were made in a temperature range from 40 to 290 K. The results are shown in Fig. 1 where only a few representative curves are plotted out of a total of 24 different measurements.

The differences in these I-V curves show the so-called 'anomalous temperature dependence' [17,18] previously explained in [19] due to the variation of the energy gap of the Al$_2$O$_3$ as a function of temperature.

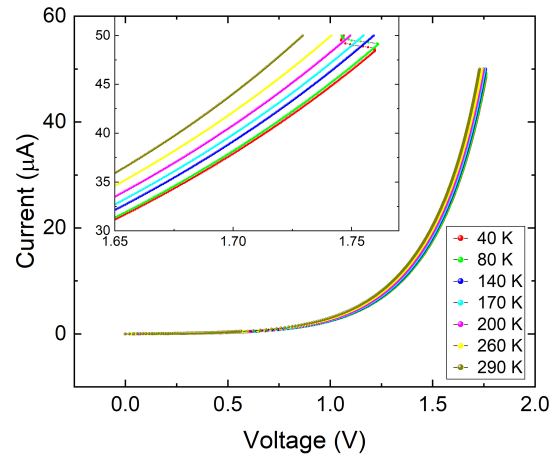

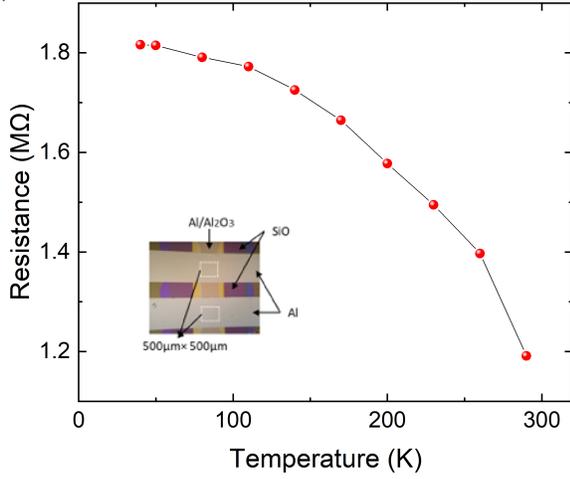

FIG. 1. I-V characteristics of Al/Al$_2$O$_3$/Al tunnel junctions at several temperatures. Inset shows zoom in view at higher voltages ~ $\phi_o/e$.

FIG. 2. Junction Resistance variation vs temperature at a bias voltage of 0.55 V. Inset shows top view of the device with two tunnel junctions. The junction area defined by SiO edges is ~ 500 μm × 500 μm (0.25 mm$^2$) as measured by optical microscopy.

The resistance vs temperature (RT) characteristic of the tunnel junction was obtained at 0.55 V bias voltage to assess the barrier quality, demonstrating an insulator-like temperature dependency of resistance (Fig. 2). More notably, a junction area resistance product $AR_N \sim 3\times10^{11} \Omega\,\mu m^2$ was calculated at high voltages. This value is significantly greater than that of earlier studies utilizing pure oxygen [20] or specialized oxidation techniques [21,22], indicating that our junctions are of higher quality.

## III. RESULTS AND DISCUSSION

Consider a particle with energy E that tunnels across a rectangular potential barrier of height $\phi_o$, as shown in the inset of figure 3. Upon applying an intermediate voltage V < $\phi_o/e$, its shape gets modified altering its tunnelling probability. Employing Simmons model [10,23], that relates the tunneling current I with the applied voltage V (see Eq. 27 in [10]), all I-V characteristics reported here were fitted for the intermediate voltage regime; $0 < V \leq \phi_o/e$. Note that in this regime the potential energy of an electron is below the barrier height value. In this voltage range, the mean barrier height is given by,

$$\bar{\phi} = \phi_o - \frac{eV}{2} \qquad (1)$$

where $\phi_o$ is the barrier height at V ~ 0.

For example, for data obtained at a temperature of 140 K, Fig. 4 (red dots) exhibits a fit using Simmons model for voltages below 1.8 V across the entire voltage range (under $\phi_o/e$).

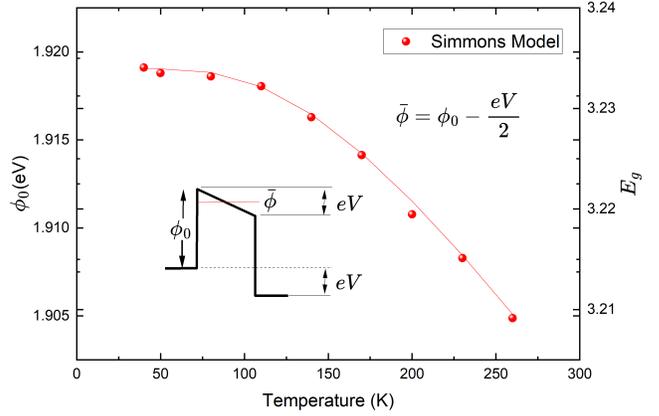

FIG. 3. Barrier height and energy gap as a function of temperature as extracted using Simmons's model, for voltages below 1.2 V (V~0 approximation). The width of the barrier is determined to be constant for temperatures from 260 K to 40 K with a value of approximately 21.45 Å. Following [19] the energy gap values as a function of temperature have been deduced from Eg(T)= γ $\phi_o$(T) relation. Additionally, a fit using the equation proposed by O'Donnell and Chen [24], is shown (red line).

The data fit yields a quadratic mean error $0.21\mu A^2$. A closer look around 0.2 V and 1.2 V shows a deviation between the fit and experiment. None of the accessible parameters in the Simmons model could be adjusted to decrease this quadratic mean error.

Consequently, the data were divided into two ranges: for voltages bellow 1.2 V, Simmons model was used, and for voltages above 1.2 V, a dissipation model that we propose was employed instead.

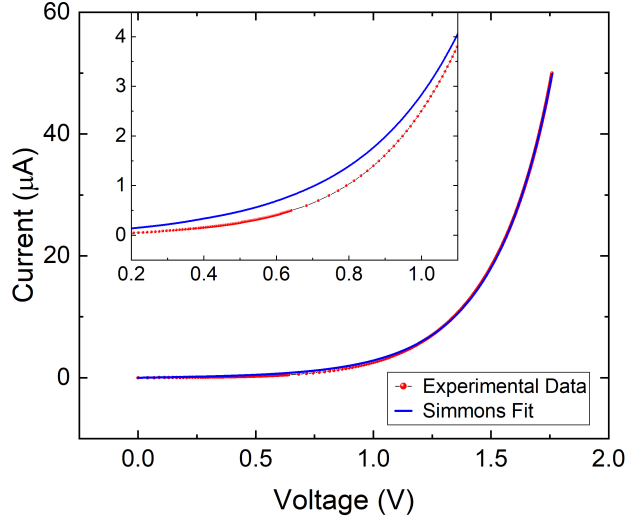

FIG. 4. Data obtained at 140 K (red dots) from our tunnel junction. The fitting of Simmons model for the complete range of voltage induces important deviations that can be appreciated by zooming in as shown in the inset of this figure.

**A. Fit for voltages below 1.2 V (Simmons model)**

Employing the Simmons model, for voltages below 1.2 V, delivered a good fit with a standard deviation lower than $4\times10^{-2}$. Here image forces were neglected as a consequence of the large dielectric constant of $Al_2O_3$ ($\kappa \geq 8$) which leads to a barrier contour that follows closely that of a rectangular barrier [10,23]. Fitting data with tunneling area A as a parameter, we find a value of approximately 0.2499 mm$^2$ for all temperatures. This value is remarkably close to the one measured by optical microscopy (500 μm × 500 μm ~ 0.25 mm$^2$) on the physical device, verifying the absence of pinholes and hotspots at the tunneling junction [25]. This demonstrates the high quality of the fabricated junction.

The I-V data were carefully fitted assuming a symmetric rectangular barrier and a free electron-effective mass. For all data analysis, we have used both forward and reverse current directions obtaining very small differences in barrier parameters. Therefore, for the rest of this work, we concentrate on the forward bias experiment.

Next, the area of the junction was fixed and the values for the width s and height $\phi_o$ of the potential barrier were found as free parameters. Results of the fitted data showed that barrier height $\phi_o$ has a decreasing behavior as the temperature increases up to the 260 K value (FIG. 3.) with a mean value of $\phi_o$ ~ 1.916 eV. While the width remained practically constant s ~ 21.45 Å across the same range of temperature.

This agrees with a previous report for high-quality tunnel junctions of similar dimensions [19] and references therein. At room temperature, 290 K, the behavior of the barrier seems to correspond to a different junction. The variation of the thickness s of the barrier also supports this hypothesis, as the thickness is practically a constant of value s ~ 21.45 Å for all temperatures below 290 K. Then, reaching 290 K it changes its behavior to a slightly reduced value of 20.83 Å. This could be explained by partial damage induced to the junction in the whole process of characterization. Indeed, given that the measurements are performed in increasing temperature order, starting at low temperature, the last characterization could reflect stress or partial damage of the junction, due to the high current densities that have tunneled through the device, at each temperature. Thus, further data analysis of energy dissipation was restricted to temperature values below 290 K to ensure the junction integrity and consequently the obtention of reliable data. In spite of the above arguments, we cannot completely rule out the possibility of a new phenomenon that sets in around this temperature which we plan to investigate in the future.

As demonstrated in our previous work [19] the energy gap Eg can be extracted from the barrier height $\phi_o$ values assuming a linear relation $E_g(T)/\phi_o(T) = \gamma$ constant at all temperatures. Given that the Eg of amorphous $Al_2O_3$ Eg(300 K) ~ 3.2 eV [26] and $\phi_o$(300 K) ~ 1.89eV (obtained from the extrapolation of our data using Simmons model) a Eg(T) = 1.69 $\phi_o$ (T) relation is found.

Additionally, from the Eg information using the theoretical equation proposed by O´Donnel and Chen [24]:
$E_g(T) = E_0 - S\langle\hbar\omega\rangle\left[\coth\left(\frac{\langle\hbar\omega\rangle}{2kT}\right) - 1\right]$ information about average phonon frequency can be extracted.
From our experimental data in this expression, we find and excellent fit (Fig 4 (solid line)) with the values Eg(0) = 3.23 ± 0.0005 eV, S = 1.52 ±0.2437, and $\hbar\omega$ = 40.2 ± 0.53 meV as can be seen in. This allows us to determine the average phonon frequency $\omega$ ~ 6 ×10$^{13}$ s$^{-1}$ which is comparable to the value obtained from speed of sound measurements [27] $\omega$ ~ 2.24×10$^{13}$ s$^{-1}$ and may be due to the different Al2O3 quality and precision of these two types of measurements. However, the same order of magnitude confirms the quality of our junctions.

**B. Dissipation model**

As discussed earlier, for voltages above 1.2 V a good fit (with the lowest standard deviation) was not possible, using Simmons model. Here a clear deviation between the experimental data and the model is observed as shown in FIG. 4. As we shall see this is due to the fact that the Simmons model does not consider energy dissipation. In the next section, we present a dissipation model that allows a good fit for voltages above 1.2 V.

In order to explain our experimental data at voltages higher than 1.2 V, we have proposed a dissipative nonlinear tunneling model.

Our previous work [9] only dealt with dissipation in the low voltage regime. This was achieved by considering a linear frictional force $f = \gamma v(x)$ where the velocity exponent is equal to one. Here we extend this model to larger voltages to encompass any loss of energy during the tunneling processes by including a friction force in the more general form of $f = \gamma \left(\frac{v(x)}{c}\right)^\alpha$ where a priori the exponent $\alpha$, the frictional coefficient $\gamma$ could take any value, while $c$ is the velocity of light.

This model incorporates a nonlinear frictional force that is proportional to an exponent of the velocity $\alpha$, in an analogous way as in complex fluid systems [28].

This force leads to the dissipation of energy as the tunneling electrons interact with the surrounding environment.
By fitting the parameters $\alpha$ and $\gamma$ to our data, we can gain a better understanding of the underlying physical mechanisms that govern the dissipative tunneling process in our system.

A particle with energy E that tunnels across a potential barrier of height $\phi_o$ can lose energy in the form of heat transferred to the potential barrier or by the release of electromagnetic radiation into the surroundings [3–5].

The total amount of energy lost, as the particles travel through the barrier over a distance $x$, due to this general frictional force is given by:

$$\Delta E(x) = \gamma \int_0^x \left(2\frac{(V(x') - E)}{m c^2}\right)^{\alpha/2} dx' \quad (2)$$

This $\Delta E(x)$ modifies the transmission coefficient for the incident particles. Calculation of the transmission coefficient in one dimensional tunneling for an electron with energy E, tunneling the barrier V(x) is usually given by means of the WKB approximation as

$$T(E) = \exp\left\{-2\int_0^s \sqrt{\frac{2m(V(x) - E)}{\hbar^2}} dx\right\} \quad (3)$$

Here $E$, corresponds to the kinetic energy component of the incident electron along the $x$ direction.
The barrier potential $V(x)$ can be simplified the following way. In the general case, we consider a symmetrical rectangular potential barrier that gets modified upon applying an intermediate voltage (i.e. V $< \phi_o/e$). As follows from the inset in figure 3, this change in shape can be accounted for by expressing the effective potential $V(x) = \varepsilon_f + \bar{\phi}$. Here $\varepsilon_f$ is the Fermi level, and the mean barrier height $\bar{\phi} = \phi_o - \frac{eV}{2}$ reflects the changes brought about by the applied voltage.

Furthermore, taking the energy dissipation into account, the transmission coefficient in (3) can now be written as,

$$T(E) = \exp\left\{-2\int_0^s \sqrt{\frac{2m(\varepsilon_f + \bar{\phi} - E'(x))}{\hbar^2}} dx\right\} \quad (4)$$

where the energy of the incident particle has been modified from $E \to E'(x) = E - \Delta E(x)$.
The final expression for our modified transmission coefficient turns out to be:

$$T(E) = \exp\left\{-4k_0 \frac{\left[(\lambda s + 1)^{\frac{3}{2}} - 1\right]}{3\lambda}\right\} \quad (5)$$

where $k_0 = \sqrt{2m(\varepsilon_f + \bar{\phi} - E)}/\hbar$, and $\lambda = \gamma (2/mc^2)^{\frac{\alpha}{2}} (\varepsilon_f + \bar{\phi} - E)^{\alpha-2}/2$.

Finally, the current density (J = I/A) expression can be calculated to obtain the junction J-V characteristic as is made by Simmons with the following equation:

$$J(V,T) = \frac{4\pi mekT}{h^3} \int_0^\infty \ln\left\{\frac{1 + \exp\left[\frac{\varepsilon_f - E}{kT}\right]}{1 + \exp\left[\frac{\varepsilon_f - E - eV}{kT}\right]}\right\} \times T(E) dE \quad (6)$$

with,
$m$= electron mass
$e$= electron charge
$k$= Boltzmann constant
$T$= temperature in K
$h$= Planck's constant
$\varepsilon_f$ = Fermi energy

**C. Fit for voltages above 1.2 V (Dissipation Model)**

In principle, the dissipation model includes four free parameters for optimization: $\phi_o$, s, α, and γ. The values of $\phi_o$ and s can be determined using the Simmons model for low voltages (V < 1.2 V), which leaves α and γ as the only free parameters. For intermediate voltages greater than 1.2 V, the dissipative model is applied. Here, the values of $\phi_o$ and s, as obtained from Simmons's model, are held constant at each temperature, and the data fitting is performed with γ and α as the adjustable free parameters.

The data obtained within the temperature range between 40 K and 260 K were fitted in a self-consistent way i.e. optimizing one parameter at the same time; alternating between α, and γ until the values no longer change significantly.

In the temperature range of 40 K to 260 K, the parameter γ (indicated by the red line) exhibits a relative variation of approximately 1.15%. On the other hand, the parameter α shows a relative variation of less than 0.4%, remaining essentially constant across this temperature span.

Both parameters show a slight increase as temperature rises, peaks around 137 K, and then decreases. We can compare γ order of magnitude with our previous linear model for the low voltage regime (FIG. 3.a in reference [9]). Notably, in both studies, the order of magnitude of γ remains consistent around 5 eV/Å compared to 1 eV/Å, although in the current research the parameter γ is fivefold greater. This is coherent with dissipation increasing at larger voltages. For higher temperatures however, a monotonic decrease with temperature of $\gamma$ and α can be observed.

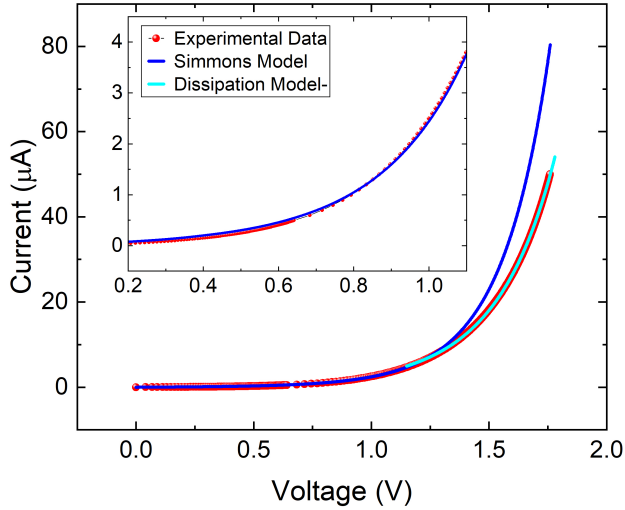

FIG. 5. Comparison of Simmons vs Dissipative model. Simmons model works well up about 1. 2 V while for higher voltages Dissipative model works better. Inset shows Simmons model fit for voltages below 1.2 V.

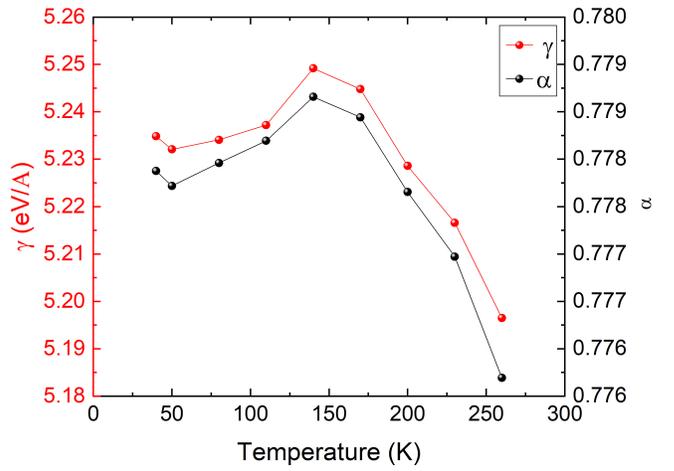

FIG. 6. γ (red) and α (black) parameters for the dissipation of energy as a function of temperature. The relative change in γ is around 1.15%. On the other hand, the relative change for α is 0.4 % which suggests a slightly weaker temperature dependence.

Indeed, as follows from Fig. 5 using Simmons Model, for voltages bellow 1. 6V, and Non-Linear Dissipative Model, for voltages above 1.6 V, an excellent fit across the full voltage range is obtained.

### D. Obtaining gamma and alpha parameters

The procedure detailed above allows finding the temperature dependence of the dissipation parameters; α, and γ, as a function of temperature depicted in Fig. 6. In this range of temperatures, the values of α are found to be less than one. In a similar fashion as the ones observed in fluid systems, sublinear drag forces describe situations in which the increase in drag force is less than the increase in velocity [28].

Lastly, using the obtained values of the parameters the dissipated energy can be calculated from the equation $\Delta E(s) = \int_0^s \gamma \left(\frac{v}{c}\right)^\alpha dx'$ where $v$ corresponds to the velocity of the incident particles.

Expressing the velocity $v$ with respect to the barrier's physical parameters can deliver a more insightful understanding of the energy loss. This is done by considering the kinetic energy $E_k = V(x) - E$ where $V(x) = \varepsilon_f + \bar{\phi}$ is the effective potential energy expressed in terms of the mean barrier height $\bar{\phi} = \phi_o - \frac{eV}{2}$ relative to the Fermi level $\varepsilon_f$, where $\bar{\phi}$ depends on the applied voltage [10,23].

By substituting the kinetic energy $E_k = \frac{1}{2} mv^2$ rearranging for $v$ and dividing by $c$, we obtain $\frac{v}{c} = \sqrt{\frac{2(\varepsilon_f + \bar{\phi} - E)}{mc^2}}$.

Consequently, the total amount of dissipated energy by the particles that cross the barrier distance *s* is then given by $\Delta E(s) = \gamma \{2(\varepsilon_f + \overline{\phi} - E)/mc^2\}^{\alpha/2} s$.

Given that the total amount of dissipated energy is expected to increase with energy of the incident particles it is more enlightening to analyze the fraction of incident energy being dissipated. This can be done by dividing this expression by E to examine the fraction of dissipated energy at the end of the barrier:

$$\frac{\Delta E(s)}{E} = \frac{\gamma}{E}\left(\frac{2}{mc^2}(\phi_0 + \varepsilon_f - eV/2 - E)\right)^{\alpha/2} s \qquad (7)$$

thereby linking the fraction of dissipated energy $\Delta E(s)/E$ directly to experimentally obtainable parameters $\phi_o$, s, V, $\gamma$ and $\alpha$.

This formulation enables the determination of the total fraction of energy loss across the barrier width, s. From this point forward we will simply refer to this as $\Delta E/E$ across the entire barrier width, omitting *s* for simplicity.

Fig. 7a illustrates the fraction of dissipated energy (Eq. 7) as a function of the energy fraction of the tunneling particle for selected values of temperature. Here the applied voltage is fixed to V = 0.95 Volts which corresponds to about half of the average barrier height for all temperatures. Note that the proportion of lost energy reduces as the particle's energy (with respect to the Fermi energy) is closer to the barrier height value. This reduction of dissipated energy fraction with energy of the incident particle agrees to the one found in our previous work [9]. Furthermore, a closer look of this dissipation as a function of temperature as shown at inset in Fig. 7a, indicates a slightly non-monotonic behavior; it decreases as the temperature increases up to a temperature around 137 K, where it rises again at higher temperatures. To investigate this dependency, the value of dissipated energy is extracted at the fraction of energy around 0.19 for each of the studied temperatures and depicted in Fig. 7b. The slight reduction, between 40 K -137 K, may be explained by the little reduction in barrier height with increasing temperature as explained in [19].

On the other hand, in the high temperature range (137 K – 260 K), $\Delta E/E$ rises with increasing temperature. The observed increase in dissipated energy fraction may be attributed to the enhanced electron-phonon interaction, which is consistent with the increase in temperature.

Lastly, doing a similar procedure by fixing the energy in equation 7, leaving the applied voltage as a free parameter the fraction of dissipated energy $\Delta E/E$ as a function of the applied voltage is given in Fig. 7c. The result shows a linear reduction in the fraction of energy with applied voltage which is consistent with Fig. 7a. This means that as the applied voltage increases so is the applied electric field to the junction which is responsible for the increase in the velocity of the incident particle. This leads to a reduction in the dissipated energy.

Finally, the energy fraction ($\Delta E/E$) as a function of barrier height is obtained for a fixed incident energy fraction $E/(\phi_0 + \varepsilon_f)$ of about 0.19 and an applied voltage of 0.94 V (which represents about half of the barrier height mean) across all temperatures is plotted in Fig. 8. Here a decrease in the energy fraction lost is observed as temperature reduces from 260 K to ~ 140 K up to barrier heights of about 1.916 eV. For higher barrier heights and lower temperatures, below 140 K up to ~ 40 K, an upturn in the fraction of dissipated energy occurs. As discuss previously, this can be explained due to the reduction of electron phonon interactions at lower temperatures, which allows to see the effect of the increase in barrier height with the reduction of temperature.

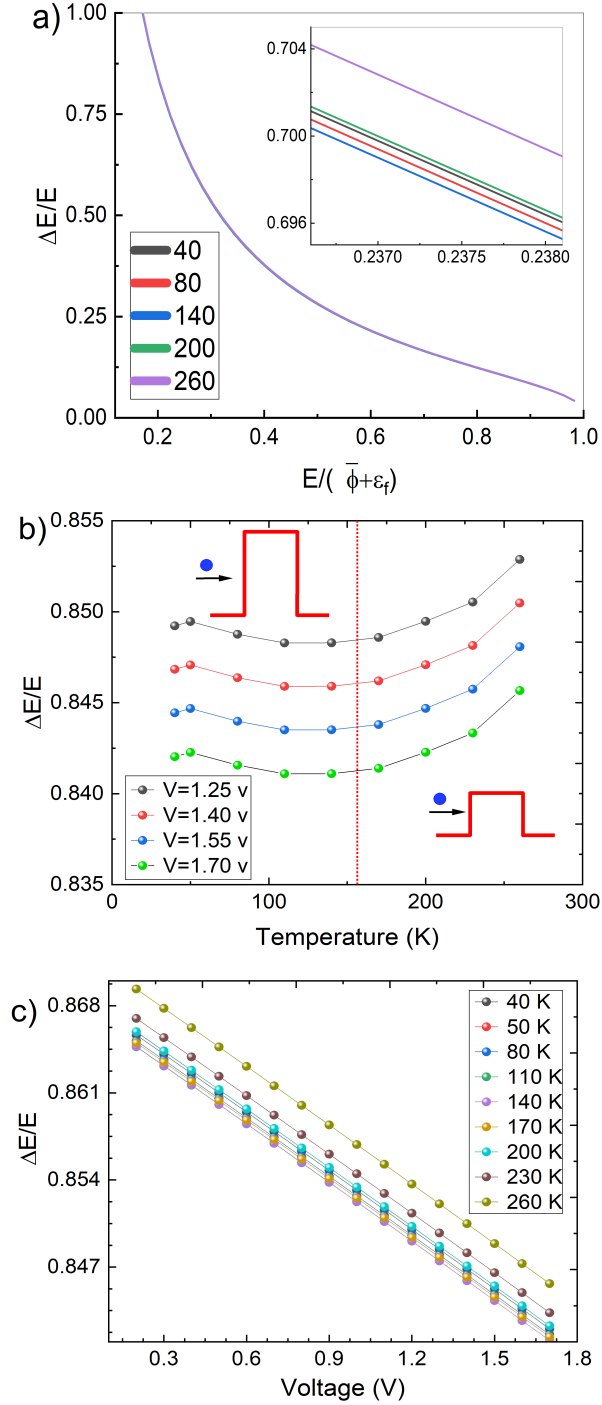

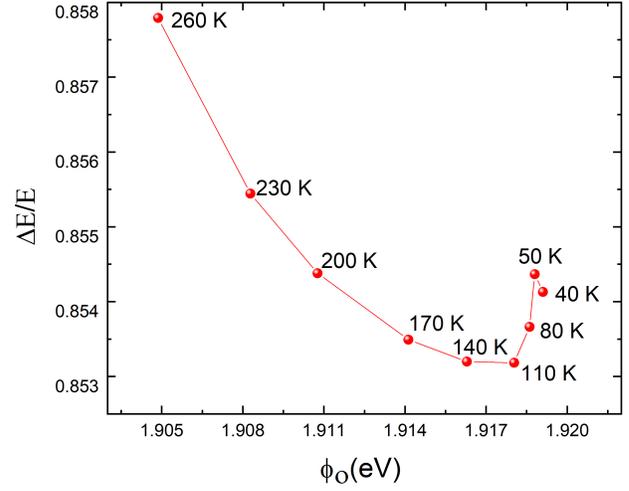

FIG. 8. Fraction of energy loss (ΔE/E) as a function of barrier height at each temperature indicated above each data point. An upturn is observed in the low temperature range around 137 K and 40 K.

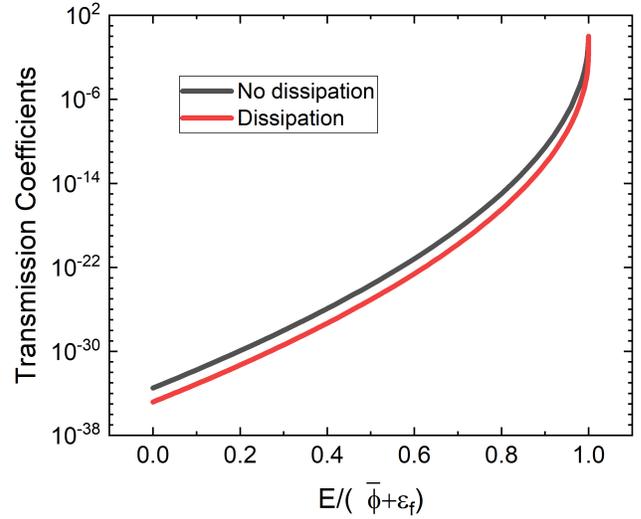

FIG. 7. Fraction of dissipated energy (ΔE/E) obtained from the non-linear dissipative model as a function of a) normalized energy $E/(\bar{\phi}+\varepsilon_f)$ b) temperature and c) applied voltage.

FIG. 9. Comparison of transmission coefficients with dissipation (Eq. 5), using the nonlinear model (red line) presented here and without dissipation (Eq. 3), using Simmon´s model (black line). The transmission coefficients obtained using the parameters V=1 Volt, $\phi_o$ = 1.897 eV, and for the case of dissipative model α = 0.776 and γ = 5.2 eV/Å.

## IV. ANALYSIS OF THE RESULTS

The previous results suggest that the dissipated energy strongly depends on the barrier height relative to the energy of the incident particle. The barrier height can be modified by two well-known mechanisms: applying a voltage or varying temperature. The effect of voltage (first mechanism) modifies the shape of the barrier, reducing the mean barrier height with linear dependency $\bar{\phi} = \phi_o - \frac{eV}{2}$, and thus lowers the effective energy barrier that the particle must overcome.

This can be explained as follows: as the electric field and voltage applied to the junction increase, the potential energy of the incident particles correspondingly rises. This in turn reduces the effective barrier height (see Eq. 1) and the tunneling probability increases due to the particle tunneling with an energy closer to the top of the barrier. Our results show this also reduces the amount of fraction of energy dissipated. Indeed, this can be observed in Fig. 7a and 7c. The increase in tunnelling probability leads to a greater number of electrons which can interact with the medium and thus greater dissipation.

Now on the other hand ¿what is the effect of the temperature? The second mechanism by which the effective barrier height lowers is by increasing the temperature. For the same arguments given above this should lead to a reduction of fraction of dissipated energy.
Indeed, measurements taken at temperatures between 40 K and about ~ 137 K, show a reduction in the fraction of dissipated energy as temperature rises (Figure 8). This suggests a lower potential barrier becomes easier for particles to overcome, consistent with the principles of quantum tunneling.

However, this trend changes at around 137 K, possibly due to the increase in electron-phonon interactions, which compete with the lowering of the barrier height and causes the increase in dissipation at high temperatures. This phenomenon distinctly defines a temperature threshold beyond which electron-phonon interactions predominate over relative barrier height and the energy of the incident particles.

Finally, inquiring into the consequences of dissipation on the tunneling probability, a comparative analysis between the Simmons and dissipative models (linear and nonlinear) were carried out. The transmission coefficients for each of the cases revealed a notable diminution in the tunneling probability upon the introduction of dissipation, as depicted in Figure 9. This outcome underlines the significant impact of dissipation on tunneling phenomena.

## V. SUMMARY

In summary, starting from experiments on tunneling junctions this investigation proposes a nonlinear dissipation model that fits well experimental results for large voltages close to the break down values. This model provides profound insights into the dynamics of energy dissipation in tunnel junctions. We underline the crucial dependency of dissipation on the relative energy of incident particles to the effective barrier height of the junction, a relationship significantly modulated by two primary mechanisms: the application of voltage and the variation of temperature.

Voltage Application: The application of voltage to the junction notably enhances the relative energy of incident particles against the barrier height, leading to a decrease in energy dissipation. This increase in applied voltage also reduces the effective barrier height, yielding a further reduction in dissipation.

Temperature Variation: Raising the temperature serves as the second mechanism for modulating the barrier height, effectively reducing dissipation up to a critical temperature of approximately 137 K. Beyond this temperature, dissipation experiences a marked increase, predominantly due to augmented electron-phonon interactions.

Moreover, the study reveals a noteworthy alteration in the behavior of dissipation parameters, γ and α, above 137 K, indicating a linear drop. This pivotal finding highlights the necessity for continued exploration at temperatures exceeding room temperature, where the characteristics of dissipation may significantly diverge.

The insights gathered from this research elucidate the important effects of applied voltage and temperature on energy dissipation within tunnel junctions. The study emphasizes the interplay between these modulatory mechanisms and their impacts on the efficiency of tunneling processes.

This research begins with an initial exploration of tunneling currents, leading to the development of a nonlinear dissipation model. This progression from empirical observations to model formulation enhances our understanding of energy dissipation mechanisms in tunnel junctions.

## AKNOWLEDGMENTS

We would like to acknowledge Cesar Talero for technical assistance and Herbert Vinck for fruitful discussions.
E. J. P wishes to thank Banco de la República (4.527); Facultad de Ciencias, Universidad de los Andes (INV-2023-162-2721); Departamento de Física, Universidad de los Andes Colombia (Conv. Financiación Equipos de Laboratorio, PROGRAMA 2021–2022); and Órgano Colegiado de Administración y Decisión (OCAD) - Ciencia, Tecnología e Innovación del Sistema General de Regalías (SGR), (BPIN 2022000100133). N.G.K. thanks la Facultad de Ciencias, Universidad de Los Andes, Colombia, for financial support through Grant No. INV-2023-162-2841.

**Author Contribution**

E. J. P. wrote approx. 60% of the manuscript, conceived, developed experiments and theory. Aided in checking and plotting the equations and figures. L.R. wrote approx. 40% of the manuscript, prepared the figures, performed the experiments, assisted in testing and developing the theory by checking and plotting the equations and figures.
N.K. conceived, tested and developed the theory. Helped in verifying and plotting the equations and figures. D.L. tested the theory, checked and plotted the equations and figures. All authors reviewed the manuscript.